# An optically modulated zero-field atomic magnetometer with suppressed spin-exchange broadening

R. Jiménez-Martínez[1,2,a)], S. Knappe[1], and J. Kitching[1]

[1]Time and Frequency Division, National Institute of Standards and Technology, 325 Broadway, Boulder, Colorado 80305, USA

[2]University of Colorado, Boulder, Colorado 80309, USA

We demonstrate an optically pumped $^{87}$Rb magnetometer in a microfabricated vapor cell based on a zero-field dispersive resonance generated by optical modulation of the $^{87}$Rb ground state energy levels. The magnetometer is operated in the spin-exchange relaxation-free regime where high magnetic field sensitivities can be achieved. This device can be useful in applications requiring array-based magnetometers where radio frequency magnetic fields can induce cross-talk among adjacent sensors or affect the source of the magnetic field being measured.

## I. INTRODUCTION

Optically pumped magnetometers are based on the precession of atomic spins about the magnetic field being measured[1]. Typically, this precession is driven by an RF field oriented either perpendicular[2] or parallel[3] to the static field axis. The presence of this RF drive field is not always desirable as it can perturb the source of the magnetic field being measured and create cross-talk among adjacent sensors in array-based applications. Optical excitation schemes[4,5], in which the optical pumping rate is modulated synchronously with the Larmor precession, can in principle overcome these difficulties. Optical excitation additionally enables remote interrogation of atomic sensors[6,7] in which the sensor head is separated by a potentially large distance from other components through the use of free-space light propagation or optical fibers. Magnetometers of this type have been investigated in a regime where the field magnitude is much larger than the width of the magnetic resonance line. Optical excitation can also be advantageous in atomic magnetometers that operate near zero field[3,8]. In some of these magnetometers[3,8] an RF field is applied to the atoms to convert the normally absorptive resonance into a dispersive resonance and reduce low-frequency noise via lock-in detection[8-13].

Recently, chip-scale versions of these magnetometers implemented with microfabricated vapor cells have been used to detect magnetic signals produced by the human brain[14] and heart[15]. The high sensitivity needed for these applications is achieved by operating the sensors in the spin-exchange relaxation-free (SERF) regime[16, 17] where broadening due to spin-exchange collisions is suppressed[18]. The SERF regime is particularly well-suited to magnetometry with miniaturized alkali vapor cells, where the alkali density must be high to provide sufficient absorption for high signal-to-noise detection of the alkali polarization[18].

As demonstrated by several authors[19, 20, 21] AC-Stark shifts, which are manifest as a fictitious field seen by the atoms[19, 20], can be advantageously used to modify the energy spectrum of alkali atoms in optical magnetometers[21] and to induce magnetic resonances[19, 20]. Here we demonstrate such an optically modulated zero-field magnetometer in a microfabricated vapor cell, which also operates in the SERF regime. We show in particular that this type of magnetometer displays the features of previous zero-field magnetometers where an RF field has been used[10-15], namely the ability to use lock-in detection to extract the device signal and to mitigate low-frequency noise. Contrary to previous magnetometers[10-15] the optical-modulation does not introduce magnetic fields that could disturb the reading of other magnetometers in the vicinity. All of these properties are critical in applications requiring multiple sensors with high magnetic sensitivity and packed in a dense array[22], such as for the detection of biomagnetic fields[22, 23], and that benefit from microfabricated magnetometers[14, 15].

## II. APPARATUS

The experimental setup shown in FIG. 1 consists of a microfabricated vapor cell with a 2 mm × 1 mm × 1 mm cavity filled with isotopically enriched $^{87}$Rb and about 1.4 amagat of $N_2$ buffer gas. The measured width (FWHM) of the optical line is $\Delta v = 25$ GHz; thus the hyperfine structures of neither the ground state nor excited state of $^{87}$Rb are resolved. The cell is placed inside a magnetic shield and heated to 170 °C by running electrical current at 350 kHz through chip resistors mounted on the cell windows. A vertical-cavity surface-emitting laser (VCSEL) provides up to 60 µW of pumping light and is detuned by approximately $\Delta v/2$ from the pressure-broadened $^{87}$Rb D1 line. The pump light is elliptically polarized so that the magnetometer signal can be extracted from the optical rotation in the polarization of the transmitted light, by means of a balanced polarimeter[11]. A circularly polarized light beam from a distributed-feedback laser provides up to 20 mW of optical power. This beam is used to shift the energy spectrum of $^{87}$Rb and hereafter is referred to as the light-shift beam. The wavelength of the light-shift beam can be tuned continually from 795.2 nm to 794.4 nm by varying the laser injection current and temperature. The pump and light-shift beams propagate along the x and y directions, respectively, as shown in FIG. 1. Their cross-sectional areas are approximately 0.64 mm². The vapor cell is tilted at 45° with respect to both x and y axes. Two Helmholtz coils and a solenoid are used to zero the magnetic field at the location of the cell. Zero-field magnetic resonances are observed by monitoring the polarimeter signal as the magnetic field $B_o$ along the y axis is scanned around zero. Dispersive resonances are detected by applying a modulating magnetic field $B_{mod}\cos(\omega_{mod}t)$, as experienced by the $^{87}$Rb atoms, along the y axis and using lock-in detection of the polarimeter signal.

### III. GENERATION OF FICTITIOUS MAGNETIC FIELDS

It is well known that the AC-Stark shift of non-resonant light on the ground state of alkali-atoms can be described in terms of fictitious electric and magnetic fields[19, 20]. These fictitious fields behave like real fields only within particular energy levels of the alkali atoms being addressed by the light. Thus they act selectively on the illuminated alkali atoms without perturbing their environment. Here we are interested in the shift that circularly polarized non-resonant light creates on the Zeeman energy levels of alkali-atoms, which can be described as the interaction of a fictitious magnetic field $B_{LS}$ with the atomic spin. This fictitious field is directed along the light propagation axis, and in the case where the hyperfine structure is not resolved, is given by[19]

$$B_{LS} = \frac{I}{h\nu_L}\frac{\sigma}{\gamma_e}\frac{x}{1+x^2}, \qquad (1)$$

where $I$ is the intensity of the light-shift beam, $h$ is Planck's constant, $\sigma$ is the on-resonance optical absorption cross-section of the pressure-broadened D1 line, $\gamma_e = 2\pi \times 28$ Hz/nT is the electron's gyromagnetic ratio, and $x = (\nu_L - \nu_o)/(\Delta\nu/2)$ is the normalized detuning of the light frequency $\nu_L$ with respect the D1 line. We characterized the vector component of the light-shift effect on the $^{87}$Rb atoms in our cell. Magnetic resonances were recorded by monitoring the polarimeter signal as the field $B_o$ was scanned around zero, then the traces were fitted to an absorptive Lorentzian from which the resonance width and offset from zero were extracted. The shift of the zero-field resonance as a function of light-shift beam intensity and detuning is shown in FIG. 2.

**IV. RESULTS**

To detect magnetic fields by use of zero-field resonances it is advantageous to have a linear response of the component of the atomic spin being probed to magnetic fields. Often, this linear response is implemented by probing the component of the spin polarization perpendicular to the orientation of both the optical pumping beam and the magnetic field. This approach requires the use of separate orthogonal light beams for pumping and probing, respectively, and often an external modulator[17] is added to modulate the optical property being monitored so that lock-in detection can be used to suppress low-frequency noise in the probe light signal. In miniaturized and portable magnetometers it has been practical to pump and probe along the same axis[12], and often with the same light beam[10,11,13-15]. In these devices the linear response of the component of the atomic spin being probed and the modulation in the probe signal are enabled by modulating the transverse magnetic field. The signal is then obtained by using lock-in detection of the probe signal at the modulation frequency $\omega_{mod}$, which yields a dispersive resonance with zero-crossing at zero field.

**A. Zero-field dispersive resonance**

In the work described here an acousto-optic modulator was used to switch on and off the intensity of the light-shift beam incident on the vapor cell, which in turn produces a time-varying fictitious magnetic field as seen by the $^{87}$Rb atoms. A zero-field dispersive resonance obtained in this manner is shown in FIG. 3(a). As a reference, a zero-field dispersive resonance obtained by use of a real RF field is shown in FIG. 3(b). In general, according to theory[3], for this type of magnetometer magnetic resonances occur when the Larmor frequency $\omega_o = \gamma_e B_o/q$ is an integer multiple of $\omega_{mod}$, where $q$ is the nuclear slowing-down factor[16]. Magnetic resonances can also be observed by referencing the lock-in detector at higher harmonics of the modulation frequency, however the zero-field dispersive resonance observed at the first harmonic has the largest slope and thus is more sensitive to magnetic field fluctuations. These resonances are sometimes referred to as parametric resonances[3,24]. In general for large detunings of the light-shift beam, where its power broadening is small compared to the width of the zero-field magnetic resonance, optically induced dispersive resonances show similar features as those produced by an RF field. For instance the zero-field resonance has a zero-crossing at zero field and finite-field resonances are observed when the Larmor frequency $\omega_o$ is an integer multiple of $\omega_{mod}$. Closely related, optically induced resonances were demonstrated previously in a system for which the atomic relaxation rate was modulated via an unpolarized light field[25].

**B. Operation in the spin-exchange relaxation-free regime**

Previous zero-field atomic magnetometers have achieved high magnetic field sensitivities by operating in the spin-exchange relaxation-free (SERF) regime[16,17], where narrow lines are obtained by suppressing spin-exchange broadening at low magnetic fields and high atomic density[18]. Our optically modulated zero-field magnetometer can also operate in this regime, as shown in FIG. 4.

The inset in FIG. 4 shows the light-induced zero-field dispersive resonance shown in Fig. 3(a) and a spin-exchange broadened resonance observed by optically driving (see Ref. 4) the spin-precession about a magnetic field of 13.6 µT. Both resonances were recorded in the same setup and at the same $^{87}$Rb atomic density, which was estimated to be $2.5 \times 10^{14}$ cm$^{-3}$ corresponding to a spin-exchange broadening of the magnetic line (FWHM) on the order of 1 µT. Clearly the zero-field resonance is much narrower due to the suppression of spin-exchange broadening in the SERF regime.

Particularly in the SERF regime and for low spin polarizations, the contribution of spin-exchange collisions to the broadening of the zero-field resonance is given, in magnetic field units, by[11,16]

$$\Delta B_{se} = \alpha \gamma_e B_{mod}^2 R_{se}^{-1}, \qquad (2)$$

where $B_{mod}$ is the amplitude of the magnetic field modulation seen by the atoms, $R_{se}$ is the spin-exchange collision rate and $\alpha$ is a constant that depends on the slowing-down factor of the atom[16]. Figure 4 shows the width of the optically induced zero-field dispersive resonance as a function of light-shift intensity. As the intensity of the light is increased, the effective time-averaged magnetic field seen by the atoms increases as well, which results in the broadening of the line due to spin-exchange collisions. In FIG. 4 one can observe a quadratic dependence of the magnetic resonance width with light intensity, the solid line corresponds to a fit of the data using Eq. (2) with an offset term. The residual width of the magnetic line is determined by other spin relaxation mechanisms[16].

## C. Suppression of low-frequency noise

An important advantage of lock-in detection is the mitigation of low-frequency noise in the device signal. Figure 5 shows the noise spectrum of the signal for magnetometers implemented by modulating the light-shift effect and a real RF field adjusted for comparable resonance with the light-shift beam. For frequencies above 2 Hz we observe that both modulation schemes yield similar magnetic-field noise levels. At lower frequencies, the higher noise levels of the magnetometer implemented with the light-shift beam could be due to the relative motion of the light-shift and pump-probe beams.

## V. DISCUSSION AND OUTLOOK

### A. Optimum off-resonance light detuning

The detuning of the light-shift beam plays an important role in the optimization of the device. As the light is tuned closer to the D1 line it broadens the magnetic resonance and degrades the magnetometer sensitivity. To minimize this effect the light should be far detuned from the optical line, so that power broadening is highly suppressed. For our experimental conditions the broadening due to non-resonant light is given by[19] $\Delta B_{light} = B_{LS}/x$, where $B_{LS}$ is given by Eq. (1) and $x > 1$ is the normalized detuning of the light. In our device $B_{LS}$ corresponds to the amplitude of the fictitious magnetic field $B_{mod}$ seen by the $^{87}$Rb atoms at the modulation frequency $\omega_{mod}$ as described above. For a given modulation frequency an optimal detuning can be estimated by setting the power broadening $\Delta B_{light}$ to be equal to the residual broadening $\Delta B_o$ due to spin-destruction collisions of Rb. The value of $B_{LS}$ is then selected to maximize the slope of the zero-field dispersive resonance with the slope approximated by the amplitude-to-width ratio of the dispersive resonance. The amplitude of the dispersive zero-field resonance is proportional to $J_0(m)J_1(m)$[3,11], where $J_0(m)$ and $J_1(m)$ are Bessel functions of the first kind and the modulation depth $m$ is defined by $m = \gamma_e B_{mod}/(q\omega_{mod})$[11]. Assuming that the broadening due to spin-exchange collisions $\Delta B_{se}$ is much smaller than $\Delta B_o$, the amplitude of the resonance is maximized when the modulation index $m \cong 1$ while its width is proportional to $\Delta B_o$[11]. Setting $B_{mod} = B_{LS}$ we find that the detuning should be $x \geq q\omega_{mod}/(\gamma_e \Delta B_o)$. The dependence of the slope of the dispersive resonance as a function of modulation frequency for $m \cong 1$ was studied in Ref. 11.

**B. Device improvements**

The implementation of the magnetometer described here in a microfabricated vapor cell and its operation in the SERF regime offers some of the features required for the realization of miniaturized and high-sensitivity magnetometers[13-15]. The current sensitivity of our magnetometer is 2 pT/$\sqrt{Hz}$, which is limited mostly by photon shot-noise and weak optical pumping. We note that previous SERF magnetometers using an RF field have achieved much better sensitivities, 20 fT/$\sqrt{Hz}$ for the magnetometer reported in Ref. 13 and 200 fT/$\sqrt{Hz}$ in Ref. 14, using microfabricated cells with probing volumes on the order of 1 mm$^3$, similar to the volume of our cell, and with optimum pumping light. We anticipate that better sensitivities can be achieved by injecting more pumping light into the cell, and that the performance of the optical modulation scheme will be similar to that obtained with a real RF field.

In addition to enhancing sensitivity there are other features of the magnetometer presented here that can be improved. Particularly, a drawback of the light-induced zero-level dispersive resonance as implemented here is that the modulation of the light-shift introduces a DC offset due to its non-zero time-averaged value. In future devices this offset can be zeroed by implementing other modulation schemes of the light-shift. For instance one could modulate the polarization of the light-shift beam from right to left circular polarization, or modulate the optical frequency of the light-shift beam from far blue-detuned to far red-detuned. An additional feature that can be explored in future devices is the use of optical fibers to guide the light from a remote location to the sensor head[6].

In summary, we have demonstrated an optically modulated zero-field atomic magnetometer operated in the SERF regime. The optical modulation approach offers all the features of previous zero-field magnetometers where an RF field has been used. The optical modulation scheme presented here, in combination with the use of laser light to heat the vapor cell[13], can enable all-optical zero-field chip-scale magnetometers, or remotely interrogated all-optical magnetometers[6,7] operating at zero fields.

**AKNOWLEDGMENTS**

The authors thank O. Alem, and K. Beloy for useful comments on the manuscript. R. J. M acknowledges support from the Roberto Rocca Education Program. This work is a contribution of NIST, an agency of the U.S. government, and is not subject to copyright.

**FIGURES**

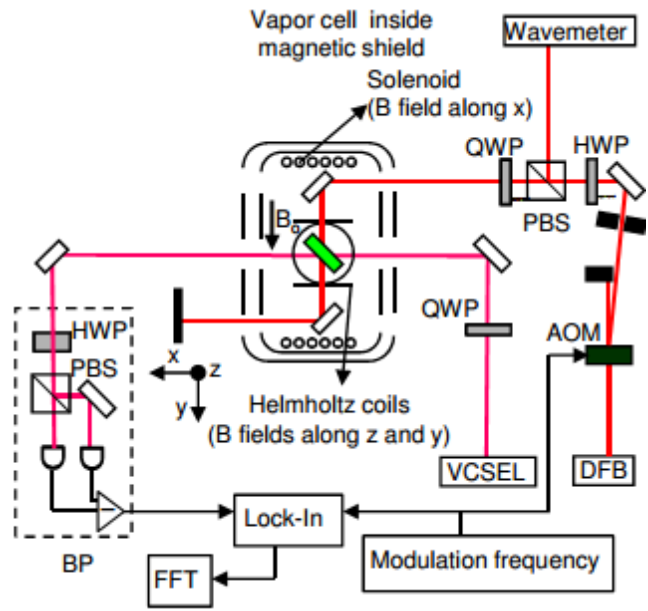

FIG. 1. Experimental setup. HWP: half-wave plate; QWP: quarter-wave plate; PBS: polarizing-beam splitter; BP: balanced photodetector; AOM: acousto-optic modulator.

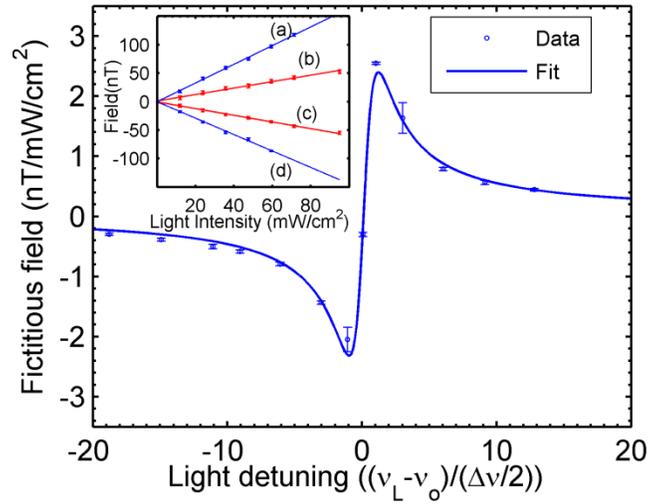

FIG. 2. Fictitious field as a function of light detuning and intensity. The data points represent the slopes of linear fits to the measured zero-field resonance shifts as a function of light intensity as shown in the inset, where (a), (b), (c), and (d) correspond to normalized detunings $x = 3$, $x = 9$, $x = -9$ and $x = -3$ respectively. The solid line represents a fit to the data based on Eq. (2). The normalized detuning of the light-shift frequency $v_L$ with respect the D1 line is defined by $x = (v_L - v_o)/(\Delta v/2)$, with $\Delta v = 25$ GHz.

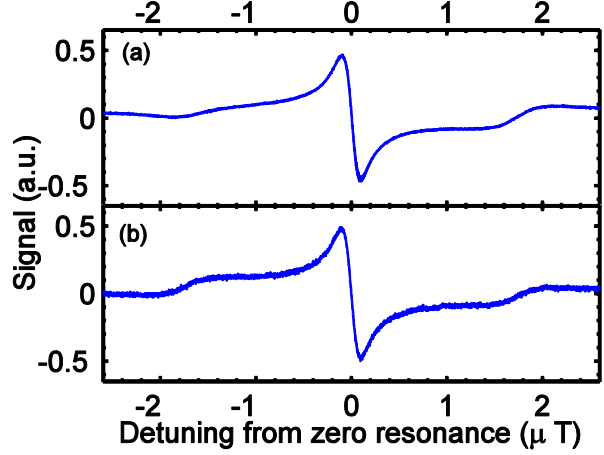

FIG. 3. Zero and finite field dispersive resonances. (a) Light-induced dispersive resonance at a light-shift beam optical power of 4 mW, which corresponds to an intensity of 600 mW/cm$^2$, and normalized detuning $x = 18.8$. At this light intensity and detuning the fictitious magnetic field is about 200 nT. (b) Dispersive resonances obtained using an RF field and having the VCSEL as the pump light while the light-shift is blocked. All resonances were obtained using lock-in at the first harmonic of the modulation frequency $\omega_{mod} = 12$ kHz, which is resonant to the Larmor frequency of the $^{87}$Rb atoms for a magnetic field $B_o \approx 1.7$ µT, and at an estimated spin-exchange collision rate $R_{se} \approx 2 \times 10^5$ s$^{-1}$. The full widths of the resonances, estimated from fits of the data to a dispersive resonance, are 182 nT and 185 nT for the resonances in (a) and (b), respectively.

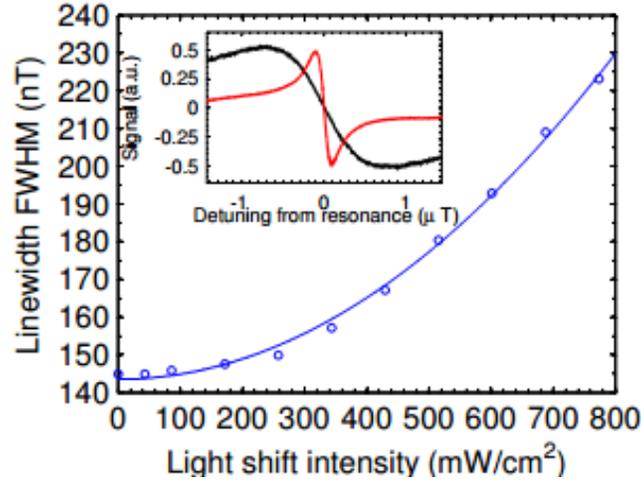

FIG. 4. Zero-field light induced dispersive resonance width as a function of light-shift intensity, at a normalized light detuning $x = 18.8$, modulation frequency $\omega_{mod} = 12$ kHz, and an estimated spin-exchange collision rate $R_{se} \approx 2 \times 10^5$ s$^{-1}$. Inset: comparison of the light-induced zero-field dispersive resonance shown in Fig. 3(a) with a spin-exchange broadened resonance observed at the same cell temperature and by optically driving (see Ref. 4) the spin-precession about a magnetic field of 13.6 µT.

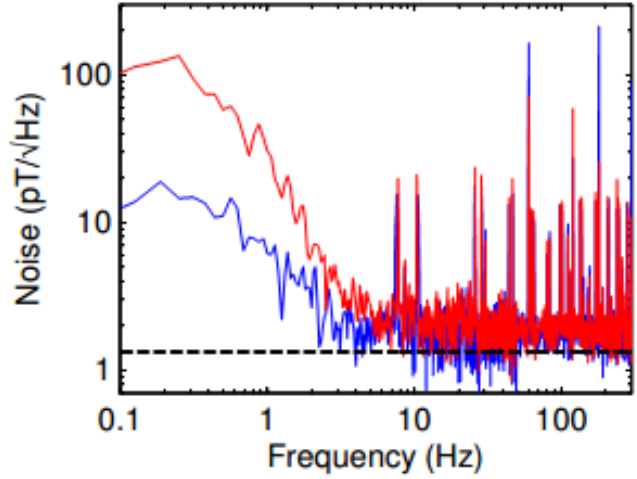

FIG. 5. Noise spectral density of the magnetometer implemented by modulating the vector AC-Stark shift (red) and using an rf field (blue) adjusted to yield a zero-field dispersive resonance with slope similar to that obtained with the light-shift beam. The detuning of the light-shift beam was $x = 18.8$. The modulation frequency $\omega_{mod} = 2$ kHz, and the estimated spin-exchange collision rate $R_{se} \approx 2 \times 10^5$ s$^{-1}$. The dashed-line corresponds to the expected shot noise due to the transmitted light levels. The bandwidth of the lock-in was 400 Hz.